\documentclass[12pt]{article}
\usepackage{amsfonts}
\usepackage{amssymb}
\begin{document}

{\LARGE
\centerline{Momentum Map and Action-Angle Variables}
\centerline{for Nambu Dynamics } }

\phantom{aaa}

\phantom{aaa}

{\large \centerline{Adnan Te\u{g}men}}Department of Physics,
University of Ankara, Faculty of Sciences, 06100 Ankara, TURKEY
\centerline{{\sl tegmen@science.ankara.edu.tr}}

\begin{abstract}
Momentum map is a reduction procedure that reduces the dimension
of a Hamiltonian system to the lower ones. It is shown that
behavior of the action-angle variables under the momentum map
generates the new action-angle variables for the reduced system
considered as a Nambu structure. The symmetrical top is given as
an illustration.
\end{abstract}
\noindent\rule{7in}{0.01in}

\section {Introduction}
It is well known that the time evolution of a Hamiltonian
dynamical system with the Hamiltonian $H$ and $n$ degrees of
freedom is governed by the equations
\begin{eqnarray}\label{basic}
\dot{q}_i =\frac{\partial (q_i,H)}{\partial (q_i,p_i)},\qquad
\dot{p}_i =\frac{\partial (p_i,H)}{\partial (q_i,p_i)},\qquad
i=1,...,n,
\end{eqnarray}
where $q_i$ and $p_i$ are the conventional phase-space
coordinates, {\it i.e.}, generalized coordinates and their
conjugate momenta, respectively. It is obvious that dimension of
the phase space of the standard Hamiltonian system is even. But in
1973 Yoichiro Nambu introduced a generalization of Hamiltonian
mechanics for arbitrary-dimensional phase spaces \cite{Nambu}. His
formalism is based on an $N$-tuple of phase space variables and
$N-1$ Hamiltonians considered as integrals of motion. He replaced
the usual Poisson bracket that is a binary operation by his own
bracket as an $N$-ary operation. In this manner, this
generalization allows us to define odd-dimensional phase spaces
besides the even ones. Indeed, if we try to determine
(\ref{basic}), for instance, for a phase space described by the
coordinates $x_1, x_2, x_3$, we should write
\begin{equation}
\dot{x}_i =\frac{\partial (x_i,H_1,H_2)}{\partial (x_1,x_2,x_3)},
\qquad i=1,2,3.
\end{equation}

Nambu focused his formalism on the three dimensional phase spaces
giving Euler equations for the free rigid body as an application.
After Nambu's proposal some authors analyzed this formalism. For
example, in Ref. \cite{Flato} Nambu dynamics is treated as
six-dimensional degenerate Hamiltonian system with three
constraints. In Ref. \cite{Mukunda} it is shown that Nambu
formalism is an embedding of three-dimensional phase space into
the four-dimensional one.

In the proceeding years there were attempts to embody the Nambu
formalism in a geometric framework. Ref. \cite{Takhtajan} treats
the formalism on a mathematical basis and also contains a
generalization of the Jacobi's identity, so-called fundamental
identity.

The aim of this work is to construct the variables in the Nambu
mechanics corresponding to the action-angle variables of Hamilton
mechanics. To do so, rather than to guess intuitively, invariance
of momentum map under canonical transformations is used. First we
give a brief summary about symplectic structures that is a
geometric description of Hamiltonian systems and their
action-angle variables.

The phase space of a dynamical system with $n$-degrees of freedom
is the cotangent bundle $T^{*}Q=M$ of the configuration space $Q$
that is the set of all possible spatial positions of objects in
the system. $M$ has a natural symplectic structure given by a
closed, nondegenerate differential $2$-form
\begin{eqnarray}
\omega^{(2)}=dq_{i} \wedge dp_{i} , \qquad i=1,...,n,
\label{2form}
\end{eqnarray}
where $\wedge$ is the exterior product. (Throughout the article
the Einstein summation rule is used for repeated indices). The
dynamics of a mechanical system is specified by a differentiable
real-valued function $H$, the Hamiltonian, defined on M, {\it
i.e.}, $H: M \rightarrow R $. The time evolution of the system is
given by the integral curve $q_{i}(t)$ of the corresponding
Hamiltonian vector field $X_H$ defined by
\begin{eqnarray}
X_H=\; ^{\sharp}dH.
\end{eqnarray}
Here the linear mapping $^{\sharp} : T^{*}M\rightarrow TM$, for
$2n$-dimensional phase space with canonical coordinates
$(q_1,p_1,...,q_n,p_n)$, is defined  by
\begin{eqnarray}
 ^\sharp dq_i={\partial}_{p_i}\;,\;\;\;\;
^\sharp dp_i=-{\partial}_{q_i}\;,\;\;\;\;
\end{eqnarray}
where TM stands for the tangent bundle of M. Then the Hamiltonian
vector field $X_H$ is explicitly
\begin{eqnarray}
X_H={\partial}_{q_{i}}H{\partial}_{p_{i}}\;-\;
{\partial}_{p_{i}}H{\partial}_{q_{i}}\;.
\end{eqnarray}
The Hamilton equations of motion are therefore obtainable easily
by
\begin{eqnarray}
\dot{x}=-^{\sharp}dH(x),
\end{eqnarray}
where $x\in \{q_1,p_1,...,q_n,p_n\}$.

The phase space arising in Hamiltonian mechanics has an additional
structure called Poisson bracket. If $f$ and $g$ are smooth, real
valued functions on $M$, the Poisson bracket $\{\;,\;\}_P$ in
canonical coordinates is defined to be
\begin{equation}
\{f,g\}_P={\partial}_{q_i}f{\partial}_{p_i}g-{\partial}_{p_i}f{\partial}_{q_i}g.
\end{equation}
The bracket operation can be defined entirely in terms of the
symplectic form without stating any particular coordinate system,
\begin{eqnarray}
\omega^{(2)}{(\;^{\sharp}df,\;^{\sharp}dH)}=\{ f,H \}_P.
\label{det}
\end{eqnarray}
Here the left hand side of (\ref{det}) is the determinant defining
the pairing of a $2$-form and the tangent bundle vector fields;
\begin{equation}
\left|
\begin{array}{rcl}
\langle \;^{\sharp}df,dq_i\rangle &
\langle\;^{\sharp}df,dp_i\rangle\\ & \\
\langle\;^{\sharp}dH,dq_i\rangle &
\langle\;^{\sharp}dH,dp_i\rangle
\end{array}
\right|,
\end{equation}
where $\langle\;,\;\rangle$ is pairing operator defined by
$\langle\;\partial_{x_i},dx_j \rangle =\delta_{ij}$. Then the
Hamiltonian equations of motion read
\begin{eqnarray}
\dot{q_i}=\omega^{(2)}{(\;^{\sharp}dq_{i},\;^{\sharp}dH)}\;\;,\;\;\;
{\dot{p}}_{i}=\omega^{(2)}{(\;^{\sharp}dp_{i},\;^{\sharp}dH)}.
\end{eqnarray}

It is well known that the canonical transformations are the
mappings $T:M\rightarrow M$ which preserve the form of Hamiltonian
equations of motion. In the symplectic context, canonical
transformations are the diffeomorphisms which preserve the
symplectic structure and are sometimes called symplectomorphisms.
The coordinate set of the action-angle variables $({\cal
I}_i,{\theta}_i)$ is a special kind of canonical transformations,
such that the angle variables $\theta_i$ appear as the cyclic
coordinates in the new system and therefore the action variables
${\cal I}_i$ are constants of motion:
\begin{equation}
\dot{\theta_i}=\partial_{{\cal I}_i}K,\qquad \dot{{\cal
I}_i}=-\partial_{\theta_i}K=0, \label{aa}
\end{equation}
where $K$ is the transformed Hamiltonian. According to the
argument stated above, the action-angle variables argue the
relation of invariance,
\begin{equation}
d{\cal I}_i\wedge d\theta_i =dq_i\wedge dp_i.
\end{equation}
Shape invariance of the Hamiltonian equations under the
transformation $(q_i,p_i)\rightarrow ({\cal I}_i,\theta_i)$
induces the invariance of the closed areas $A_i$ projected onto
the $(q_i,p_i)$ planes. Thus under this circumstance the action
variables are defined in terms of these areas:
\begin{equation}
{\cal I}_i=\frac{1}{2\pi}\oint p_idq_i=\frac{1}{2\pi}A_i.
\end{equation}
On the other hand the quantities $\dot{\theta_i}=\nu ({\cal I}_i)$
correspond to the sweeping  frequencies of the areas $A_i$.

\section{Nambu Structure and Momentum Map}
Nambu structure is a generalization of
the Hamiltonian structure, proposed by Y. Nambu in 1973
\cite{Nambu}. In Nambu's formalism the equations of motion for
$N$-dimensional phase space with coordinates $x_1,\dots ,x_n$, are
governed by $N-1$ Hamiltonians $H_1,\dots ,H_{N-1}$ and a closed,
nondegenerate $N$-form $\omega^{(N)}=dx_1\wedge \cdots \wedge
dx_N$ (volume form on Nambu phase space $M_N$) :
\begin{equation}
\dot{x_i}= \omega^{(N)}({\widetilde{dx}}_i
,{\widetilde{dH}}_1,\dots , {\widetilde{dH}}_{N-1})\;\;,\;\;\;i\in
\{1,\dots ,N\}, \label{wn}
\end{equation}
here we adopt the linear map $^{\sim } $ defined by
\begin{equation}
\widetilde{dx_i}=\partial_{x_i},
\end{equation}
so that $\widetilde{df}=\partial_{x_i}f\partial_{x_i}$. (\ref{wn})
defines the Nambu bracket $\{.,\dots ,.\}_N$
\begin{equation}
\omega^{(N)}({\widetilde{dx}}_i ,{\widetilde{dH}}_1,\dots ,
{\widetilde{dH}}_{N-1})=\{x_i ,H_1, \dots ,H_{N-1}\}_N,
\end{equation}
which is given explicitly by the Jacobian
\begin{equation}
\frac{\partial {(x_{i},{H}_{1},\dots ,{H}_{N-1})}} {\partial
{(x_{1},\dots x_{N})}}=\{x_i ,H_1, \dots ,H_{N-1}\}_N, \label{nb}
\end{equation}
that is the generalization of the usual Poisson bracket just as
stated originally by Nambu. It is remarkable to point out that the
systems with $n$-degrees of freedom ($n\geq 2$), {\it i.e.}, even
dimensional phase space ($N=2n$) with the canonical coordinates
$(q_1,p_1,\dots ,q_n,p_n)$, can be treated in two ways. First, it
can be considered as a Hamiltonian system with the usual Poisson
bracket. Second, it can be interpreted as a Nambu system with the
bracket
\begin{equation}
\dot{f}=\frac{\partial (f,H_1,\dots ,H_{N-1})}{\partial
(q_1,p_1,\dots ,q_n,p_n)}.
\end{equation}
In this case there appears an inevitable normalization function
that is also a constant of motion in front of the right hand side
of (\ref{nb}) \cite{Chatterjee, Zachos, Tegmen}. In order to get
the correct equations of motion the Nambu bracket must then be
normalized properly. Since this approach does not destroy the
fundamental ideas within this paper, all Nambu brackets will be
considered as normalized.

On the other hand when considering the odd dimensions the
phase-space coordinates are not the canonical ones in the Poisson
sense. The systems with the non-canonical coordinates can be
defined as the substructure of a standard Hamiltonian system
\cite{Marsden-Ratiu}. A general feature for Hamiltonian systems
with symmetry, for instance integrable or superintegrable, is that
it is possible to reduce the dimension of the system to the lower
ones. This reduction procedure is achieved by appealing to
non-canonical coordinates and may be used for non-integrable
systems, just as well as  for the integrable or superintegrable
ones. The reduced phase space gets a new non-canonical bracket
from the original phase space under this reduction since the
invariance of the equations of motion is desired. Momentum map is
a reduction procedure corresponding to a particular conserved
quantity. Consider, for example, the system for the free rigid
body. It is well known that the magnitude of the angular momentum
vector is a constant of motion for this system. The momentum map,
consisting of the angular momentum is given by
\begin{eqnarray}
L_{1}&=&\frac{P_{\theta}\sin \theta \cos \psi +P_{\varphi}\sin
\psi - P_{\psi}\cos \theta \sin \psi }{\sin \theta}, \nonumber \\
L_{2}&=&\frac{P_{\varphi}\cos \psi -P_{\theta}\sin \theta \sin
\psi -P_{\psi}\cos \theta \cos \psi }{\sin \theta}, \label{rb}\\
L_{3}&=&P_\psi .\nonumber
\end{eqnarray}

It is clear that this map reduces the six-dimensional phase space
to the three dimensional one with the non-canonical coordinates
$L_1, L_2, L_3$. The time evolution of the momentum map follows
the Euler equations:
\begin{equation}
{\dot{L}}_{1}={L_2}\frac{L_3}{I_3}-{L_3}\frac{L_2}{I_2},\quad
{\dot{L}}_{2}={L_3}\frac{L_1}{I_1}-{L_1}\frac{L_3}{I_3},\quad
{\dot{L}}_{3}={L_1}\frac{L_2}{I_2}-{L_2}\frac{L_1}{I_1},\label{eu}
\end{equation}
where $I_{j}\;,\;(j=1,2,3)$ are the inertia momenta with respect
to the principal axes. The map (\ref{rb}) defines a Nambu
structure with the 3-form $\omega^{(3)}=dL_{1}\wedge dL_{2}\wedge
dL_{3}$ such that
\begin{eqnarray}
{\dot{L}}_{j}=
\omega^{(3)}{(\widetilde{dL}_j,\widetilde{dH}_1,\widetilde{dH}_2)},
\end{eqnarray}
where $H_1=\frac{1}{2}{({L_1}^{2}+{L_2}^{2}+{L_3}^{2})}$ is the
reduced phase space, {\it i.e.}, the invariant sphere, and
$H_2=\frac{1}{2}\left({L_1}^2/{I_1}+{L_2}^2/{I_2}+{L_3}^2/{I_3}\right)$
is the Hamiltonian for the reduced system. The trajectory $L(t)$
on the reduced phase space is the intersection of an ellipsoid and
a sphere. The reduction operation mentioned above can be
summarized in a differential geometric point of view as the
following. (From now on, for the sake of clarity it will be
concentrated on three-dimensional phase spaces. It is of course
possible to extend it to higher dimensions using similar
arguments).

Given a symplectic structure described by the 2-form in
(\ref{2form}) and the Hamiltonian $H_0=H_0 (q_i,p_i)$, the
transition to the Nambu structure is determined by the
transformation
\begin{equation}
x_{j}=x_{j}{(q_{i},p_{i})},
\end{equation}
comprising the components of a momentum map. This transformation
must obey the conservation condition
\begin{equation}
\dot{x_j}=\omega^{(2)}(^\sharp dx_j,^\sharp dH_0)=
\omega^{(3)}(\widetilde{dx}_j,\widetilde{dH}_1,\widetilde{dH}_2).
\label{cc}
\end{equation}
If the reduced Hamiltonian is chosen as $H_1=H_1(x_1,x_2,x_3)$,
then the condition (\ref{cc}) shows that the transition functions
satisfy the commutation relations
\begin{equation}
\{x_j,x_k
\}_P=\epsilon_{jkl}\partial_{x_l}H_2,\;\;,\;\;\,(j,k,l=1,2,3),\label{com}
\end{equation}
determining what the second Hamiltonian $H_2$ must be. In the
rigid-body example stated above, the commutation relations
(\ref{com}) correspond to the angular momentum algebra and the
calculation of the other Hamiltonian $\frac{1}{2}L^2$ is
straightforward.

Clearly for the transformations $\dot{x_j}=0$, such as the Hopf
fibration (that will be considered later), the second Hamiltonian
$H_2$ is of the form $H_2=H_2(H_1)$ ,{\it i.e.}, any function of
$H_1$. On the other hand the action of the momentum map on the
symplectic 2-form $\omega^{(2)}$ can be represented by an
operation such as $\hat{\cal D}\omega^{(2)}=\omega^{(3)}$ defined
by
\begin{equation}
\hat{\cal D}\omega^{(2)}=\imath_{X_{x_1}}\omega^{(2)}\wedge
\imath_{X_{x_2}}\omega^{(2)}\wedge \imath_{X_{x_3}}\omega^{(2)}
=dx_1\wedge dx_2\wedge dx_3,
\end{equation}
where $\imath$ is the usual interior product. For the 2-form
$\omega^{(2)}=dq\wedge dp$ it is obvious that $\hat{\cal
D}\omega^{(2)}=0$. The invariance under the canonical
transformations is the final remark about the momentum map. If
$y_1,y_2,y_3$ are the transformed coordinates with respect to the
new canonical coordinates then $dx_1\wedge dx_2\wedge
dx_3=dy_1\wedge dy_2 \wedge dy_3$. This fact will be the main idea
throughout the text.

\section{Hopf Fibration and Behavior of Action-Angle Variables}

In this section, Hopf fibration is given as model example for the
momentum map and it is shown how to transform the action-angle
variables under this reduction process.

It is well known that, in a two-dimensional phase space, the most
suitable choice as canonical transformation for the action-angle
variables is the set of polar coordinates $(r,\theta)$, just as
the spherical coordinates for four-dimensional phase space shown
in the following. Thus the question arises naturally: Is the set
of spherical coordinates a suitable choice for the action-angle
variables for a three-dimensional phase space? This section shows
that the answer to this question is affirmative. The question may
seem a bit trivial but it is meaningful when considering the
bracket of the system, because such a choice must not effect the
structure of the Nambu bracket. Next section confirms this fact.

Consider the 4-dimensional harmonic oscillator with the
Hamiltonian
\begin{equation}
H_0=\frac{1}{2}{({p_1}^{2}+{p_2}^{2}+{q_1}^{2}+{q_2}^{2})},
\end{equation}
describing  a 3-sphere $S^3$. The Hopf fibration is given by the
momentum-map components
\begin{equation}
x_{1}=2{(q_{1}q_{2}+p_{1}p_{2})}, \quad
x_{2}=2{(q_{2}p_{1}-q_{1}p_{2})},  \quad
x_{3}={q_1}^{2}+{p_1}^{2}-{q_2}^{2}-{p_2}^{2},
\end{equation}
obeying ${\dot{x}}_{j}=0$. Such a map converts the three-sphere
$S^3$ into the two-sphere $S^2$, identified by the reduced
Hamiltonian
\begin{equation}
H_1=\frac{1}{2}{({x_1}^{2}+{x_2}^{2}+{x_3}^{2})}^{1/2}
\end{equation}

Though a different set of spherical coordinates can be used for a
four-dimensional sphere to define a canonical transformation, we
prefer the following one used for the symmetrical top,
\begin{eqnarray}
q_{1}&=&r^{1/2}\cos (\frac{\theta}{2})\cos
[\frac{1}{2}{(\alpha+\beta)}],\qquad
p_{1}=r^{1/2}\cos (\frac{\theta}{2})\sin [\frac{1}{2}{(\alpha +\beta)}], \\
q_{2}&=&r^{1/2}\sin (\frac{\theta}{2})\cos
[\frac{1}{2}{(\alpha-\beta)}], \qquad p_{2}=r^{1/2}\sin
(\frac{\theta}{2})\sin [\frac{1}{2}{(\alpha-\beta)}],
\end{eqnarray}
where $r={p_1}^{2}+{p_2}^{2}+{q_1}^{2}+{q_2}^{2}$ and $\alpha \in
[0,2\pi )$, $\theta \in [0,\pi )$, $\beta \in [0,4\pi )$. Now if
we let
\begin{eqnarray}
r_1=r^{1/2}\cos (\frac{\theta }{2}),\quad
{\theta}_{1}=\frac{1}{2}{(\alpha +\beta )},\quad r_2=r^{1/2}\sin
(\frac{\theta }{2}),\quad {\theta}_{2} =\frac{1}{2}{(\alpha -\beta
)},
\end{eqnarray}
we get the closed orbits in the $(q_i,p_i)$-planes:
\begin{equation}
q_1=r_{1}\cos {\theta }_{1},\quad
p_1=r_{1}\sin {\theta}_{1},\quad
q_2=r_{2}\cos {\theta }_{2},\quad
p_{2}=r_{2}\sin {\theta }_{2},
\end{equation}
where ${r_1}^{2}+{r_2}^{2}=r$. Consequently, the symplectic
two-form $\omega^{(2)}=dq_1\wedge dp_1+dq_2\wedge dp_2$ transforms
to
\begin{equation}
\omega^{(2)}=r_{1}dr_{1}\wedge d\theta_{1}+ r_{2}dr_{2}\wedge
d\theta_{2}.
\end{equation}
Obviously, this transformation describes the invariant action-angle variables
\begin{equation}
\omega^{(2)}=d{\cal I}_{1}\wedge d{\theta}_{1}+ d{\cal
I}_{2}\wedge d{\theta}_{2}
\end{equation}
with the action variables ${\cal I}_{1}={r_1}^{2}/2$, ${\cal
I}_{2}={r_2}^{2}/2$.

On the other hand, the transition to the Nambu $3$-form
$\omega^{(3)}$ is possible via the transition function
$F=-8{H_0}^2$ such as
\begin{equation}
\omega^{(3)}=dF\wedge \omega^{(2)},
\end{equation}
or in an explicit form
\begin{equation}
dx_1\wedge dx_2\wedge dx_3=8r_1r_2({r_1}^2+{r_2}^2) dr_1\wedge
dr_2\wedge d(\theta_1-\theta_2). \label{8r}
\end{equation}
When written in terms of the action-angle variables, (\ref{8r})
follows
\begin{equation}
dx_1\wedge dx_2\wedge dx_3=16{({\cal I}_1+{\cal I}_2)}d{\cal
I}_1\wedge d{\cal I}_2\wedge d(\theta_1-\theta_2),
\end{equation}
which has the more compact form
\begin{equation}
\omega^{(3)}=dG\wedge (d{\cal I}_i\wedge d\theta_i),
\end{equation}
where $G=-8({\cal I}_1+{\cal I}_2)^2$. If one defines $\varphi
=\theta_1-\theta_2$ in (\ref{8r}), then
\begin{equation}
dx_1\wedge dx_2\wedge dx_3=r^2\sin \theta dr\wedge d\theta \wedge
d\varphi . \label{fin}
\end{equation}
At this stage it is not difficult to define the new action-angle
variables. Indeed the rearrangement of (\ref{fin}) as the
following
\begin{equation}
\omega^{(3)}=d(r^3/3)\wedge d(-\cos \theta)\wedge d\varphi=d{\cal
J}\wedge d\mu \wedge d\varphi , \label{mu}
\end{equation}
gives the action variable in terms of the volume $V$ enclosed in
the phase space;
\begin{equation}
{\cal J}=\frac{1}{4\pi }V=\frac{1}{8\pi }\;\;\epsilon_{klm}\int
x_kdx_l\wedge dx_m , \qquad k,l,m =1,2,3.
\end{equation}
Note that ${\cal J}$ does not have the dimensions of an angular
momentum. Therefore we should keep this in mind when we talk about
"action". Perhaps the best way to emphasize the action-angle
variables is to say "area-angle" variables for the symplectic
structures and "volume-solid angle" variables for the Nambu
structures. Indeed, if we consider the solid angle element
$d\Omega =\sin \theta d\theta \wedge d\varphi$, the Nambu
three-form can be expressed as
\begin{equation}
\omega^{(3)}=d{\cal J}\wedge d\Omega .
\end{equation}

\section{Free Symmetrical Top}

In this final section, the illustration of the argument is given
by the example of free symmetrical top. We show that the Nambu
bracket gives the correct equations of motion when expressed in
the new variables ${\cal J}, \mu ,\varphi$ in (\ref{mu}). We shall
seek a set of equations of motion similar to (\ref{aa}),
\begin{equation}
\dot{{\cal J}}=\{ {\cal J}, K_1, K_2 \}_N =0 ,\quad
\dot{\mu}=\{\mu, K_1, K_2 \}_N , \quad
\dot{\varphi}=\{\varphi,K_1, K_2 \}_N ,\label{mudot}
\end{equation}
where $K_1=K_1({\cal J},\mu ,\phi )$ and $K_2=K_2({\cal J},\mu
,\phi )$ are the transformed forms of $H_1$ and $H_2$
respectively. The first equation is a direct consequence of the
Liouville theorem corresponding to the invariance of the volume in
the phase space.

A free symmetrical top is a free rigid body with $I_1=I_2$. Thus
the Euler equations of motion (\ref{eu}) reduce to
\begin{equation}
\dot{L}_1 = L_3 \left( \frac{1}{I_3}-\frac{1}{I_1}\right)
L_2=\omega L_2, \qquad
\dot{L}_2=-\omega L_1 ,\qquad
\dot{L}_3=0.
\end{equation}
The solution of these equations states that the vector ${\bf L}$
rotates uniformly about the vertical axis of the body, say $z$,
with the constant frequency $\omega $. Also the angle $\theta $
between the $z$-axis and the vector ${\bf L}$ is constant. This
clearly implies $\dot{\mu }=0$.

Since the phase space of the rigid body is identified with the
angular-momentum sphere, it is easy to construct the action
variable such as ${\cal J}=(2H_2)^{3/2}/3$, which induces
\begin{eqnarray}
K_1&=&\frac{(3{\cal J})^{2/3}}{2}\left[ \frac{1}{I_1}-\mu^2 \left(
\frac{1}{I_1}-\frac{1}{I_3}\right) \right] , \\
K_2&=&(3{\cal J})^{2/3}/2.
\end{eqnarray}
We are now in a position to perform the brackets in (\ref{mudot}).
If we write the brackets explicitly, it is clear that
\begin{eqnarray}
\dot{{\cal J}}&=&\partial_{\mu
}K_1\partial_{\varphi}K_2-\partial_{\varphi}K_1
\partial_{\mu}K_2 =0, \\
\dot{\mu}&=&\partial_{\varphi }K_1\partial_{\cal
J}K_2-\partial_{\cal J}K_1\partial_{\varphi}K_2=0, \\
\dot{\varphi}&=&\partial_{{\cal J}}K_1
\partial_{\mu
}K_2-\partial_{\mu}K_1 \partial_{\cal J}K_2 = L_3 \left(
\frac{1}{I_3}-\frac{1}{I_1}\right) . \label{phidot}
\end{eqnarray}
Thus the immediate solution of (\ref{phidot}) is $\varphi =\omega
t+$ constant. This is in accordance with the well known results
about the symmetrical top.

\phantom{AA}

\noindent{\Large{\bf Acknowledgments}}\newline

The author wishes to express his appreciations to
T. Dereli, A. Y{\i}lmazer and A. Ver\c{c}in for their helpful
comments and careful reading. This work was supported in part by
the Scientific and Technical Research Council of Turkey
(T\"{U}B\.{I}TAK).

\end{document}